\documentclass{achemso}
\usepackage[version=3]{mhchem} % Formula subscripts using \ce{}
\usepackage[T1]{fontenc}       % Use modern font encodings
\usepackage{graphicx}					% grafiken einbinden
\usepackage{epstopdf}					%eps grafiken einbinden
\usepackage{amsmath}
\usepackage{color}

\setkeys{acs}{usetitle = true}

\title{Determination of Energetic Positions of Electronic States and the Exciton Dynamics in a $\pi$-Expanded \emph{N}-Heterotriangulene Derivative Adsorbed on Au(111)}

\author{Jakob Steidel}
\affiliation{Ruprecht-Karls-Universit{\"a}t Heidelberg, Physikalisch-Chemisches Institut, Im Neuenheimer Feld 253, 69120 Heidelberg, Germany}
\author{Ina Michalsky}
\affiliation{Ruprecht-Karls-Universit{\"a}t Heidelberg, Organisch-Chemisches Institut, Im Neuenheimer Feld 270, 69120 Heidelberg, Germany}
\author{Mohsen Ajdari}
\affiliation{Ruprecht-Karls-Universit{\"a}t Heidelberg, Physikalisch-Chemisches Institut, Im Neuenheimer Feld 253, 69120 Heidelberg, Germany}
\author{Milan Kivala}
\affiliation{Ruprecht-Karls-Universit{\"a}t Heidelberg, Organisch-Chemisches Institut, Im Neuenheimer Feld 270, 69120 Heidelberg, Germany}
\author{Petra Tegeder}
\email {tegeder@uni-heidelberg.de}
\affiliation{Ruprecht-Karls-Universit{\"a}t Heidelberg, Physikalisch-Chemisches Institut, Im Neuenheimer Feld 253, 69120 Heidelberg, Germany}
\email{tegeder@uni-heidelberg.de}
\phone{+49 (0) 6221 548475}

\clearpage
\newpage

\begin{document}
\clearpage
\newpage

\begin{abstract}
Bridged triarylamines, so-called N-heterotriangulenes (N-HTAs) are promising organic semiconductors for applications in optoelectronic devices. Thereby the electronic structure at  organic/metal interfaces and within thin films as well as the electronically excited states dynamics after optical excitation is essential for the performance of organic-molecule-based devices. Here, we investigated the energy level alignment and the excited state dynamics of a N-HTA derivative adsorbed on Au(111) by means of energy- and time-resolved two-photon photoemission spectroscopy. We quantitatively determined the energetic positions of several occupied and unoccupied molecular (transport levels) and excitonic states (optical gap) in detail. A transport gap of 3.20 eV and an optical gap of 2.58 eV is determined, resulting in an exciton binding energy of 0.62 eV. With the first time-resolved investigation on a N-HTA compound we gained insights into the exciton dynamics and resolved processes on the femtosecond to picosecond timescale.

\end{abstract}

\begin{figure*}
\centering
\resizebox{0.7\hsize}{!}{\includegraphics{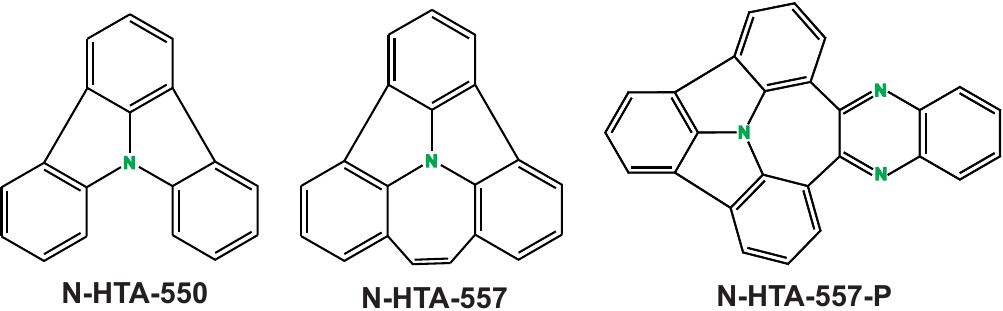}}
\caption{The N-heterotriangulenes N-HTA-550 and N-HTA-557. The $\pi$-expanded derivative N-HTA-557-P investigated in the present study.}
\label{N-HTA-557-P}
\end{figure*}

N-heteropolycyclic aromatic compounds represent a promising molecule class for applications in functional
organic materials, since their electronic structure and the resulting individual molecular properties
are efficiently tuneable by number and position of nitrogen atoms in the aromatic structural backbone.
The isosteric replacement of a C–H unit by N leaves the geometric structure unchanged, while the electronic structure are altered. For instance, they are promising candidates for electron-transporting (n-channel) semiconductors, which are of great interest for organic field effect transistors \cite{klauk2007, wurthner2011, Bunz2013, miao2014, bunz2015}. N-heterotriangulenes (N-HTAs), in which the originally propeller-shaped triphenylamine unit is locked into a planar configuration via bridging with appropriate molecular moieties (e.g. carbonyl or dimethylmethylene) \cite{Hammer2015, Michalsky2022, Wagner2022} is a class of functional molecules with high potential for optoelectronic materials \cite{Meinhardt2016, Hirai2019, Schaub2020, Krug2020}. In electron donor/acceptor systems N-HTAs act as the donor compound \cite{Kader2019}.
Recently, we analyzed the electronic and absorption properties of two N-HTA derivatives (N-HTA-550 and N-HTA-557, see Fig. \ref{N-HTA-557-P}) at the interface to Au(111) and within thin molecular films using vibrational and electronic high resolution electron energy loss spectroscopy and  quantum chemical calculations. Inter alia, we found that the additional -C=C- bridge forming the 7-membered ring in N-HTA-557 resulted in a pronounced reduction of the optical gap size by 0.9 eV from 3.4 eV in N-HTA-550 to 2.5 eV in N-HTA-557 due to an increase of the $\pi$-conjugated electron system \cite{Ajdari2021b}. Thus, using structural extensions or substitution patterns opens the opportunity for fine-tuning the electronic properties \cite{Hirai2019, Deng2019, Michalsky2022}. Here, we investigated the electronic structure and electronically excited state dynamics of a novel $\pi$-expanded N-heterotriangulene derivative (N-HTA-557-P, see Fig. \ref{N-HTA-557-P}), which consists of an
electron-deficient pyrazine ring (electron accepting unit)\cite{Michalsky2022} adsorbed on Au(111) utilizing two-photon photoemission (2PPE) spectroscopy.
2PPE  has been proven to be a powerful tool, which
provides the unique opportunity for a quantitative determination of the energetic position of occupied as well as unoccupied
molecular electronic states (transport levels) but also excitonic
states (optical gaps) and therefore exciton binding energies \cite{Zhu2004, Zhu2004a, Lindstrom2006, Muntwiler2010, bronner2012influence, bronner2011switching, Armbrust2012, bronner2013, Bogner2016, Bogner2015, Stein2017, Gerbert2017, Gerbert2017a, Bronsch2019, Stein2019, Ajdari2020, Stein2021, Stein2021b, Stallberg2022}. In addition, femtosecond (fs) time-resolved 2PPE allows to gain insights into the electronically excited states dynamics after optical excitation such as exciton formation and decay dynamics.

Here we thoroughly elucidated the electronic structure of N-HTA-557-P/Au(111) in detail, including the determination of transport levels, the optical gap and accordingly the exciton binding energy. In particular for applications of this material in organic field effect transistors or solar cells knowledge about  the energetic positions of transport levels (affinity levels and ionization potentials) are extremely important. Moreover, fs-time-resolved 2PPE, allowed to gain insights into the exciton dynamics. Our study represents the first time-resolved investigation on a N-HTA-derived compound, which is of great interest with respect to applications of this donor material in organic solar cells.

\begin{figure*}[h!]
\centering
\resizebox{0.95\hsize}{!}{\includegraphics{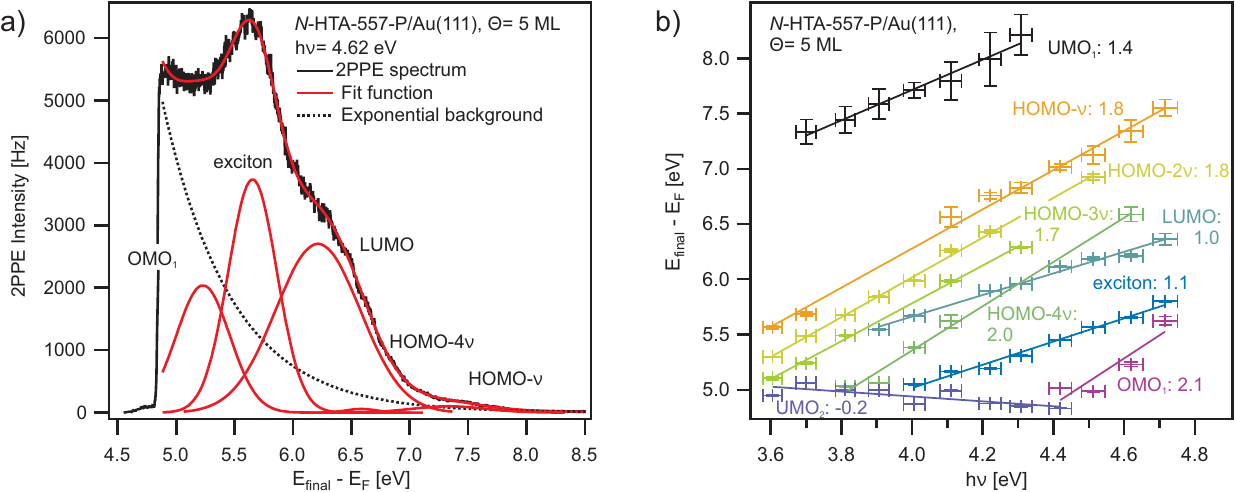}}
\caption{a) 2PPE spectrum of 5 monolayer (ML) N-HTA-557-P adsorbed on Au(111). The data are fitted with an exponential background and Gaussian-shaped peaks (red curves). The energy axis reveals the final state ($E_{Final}$) of photoemitted electrons with respect to the Fermi energy $E_{F}$ ($E_{Final} - E_{F} = E_{kin} + \Phi$); thus, the low-energy cutoff corresponds to the work function ($\Phi$) of the adsorbate/substrate system. b) Photon-energy-dependent peak position extracted to assign peaks observed in the 2PPE spectrum to occupied, unoccupied intermediate or final electronic states. A slope of 1 suggests that a peak originates from an unoccupied intermediate state, a slope of zero from an unoccupied final state (located above the vacuum level), while a slope of 2 is related to peaks originating from occupied states. The slopes from the fits are given next to the respective data.}
\label{2PPE_data}
\end{figure*}

To gain insight into the energy level alignment, i.e., the energetic positions of adsorbate-derived occupied  as well as unoccupied electronic states we conducted photon-energy-dependent 2PPE measurements \cite{Bronner2012a, bronner2013, Bogner2015, Bogner2016, Stein2017, Stein2019, Ajdari2020, Stein2021, Stein2021b}.
Figure \ref{2PPE_data}a shows an exemplarily 2PPE spectrum of 5 monolayer (ML) N-HTA-557-P adsorbed on Au(111) recorded with $h \nu$ = 4.62 eV (for additional 2PPE data see supporting information). Several photoemission features could be detected. On the basis of their photon-energy-dependency they are assigned to occupied and unoccupied molecular orbitals as well as an excitonic state (see Fig. \ref{2PPE_data}b).

\begin{figure}[h!]
\centering
\resizebox{0.45\hsize}{!}{\includegraphics{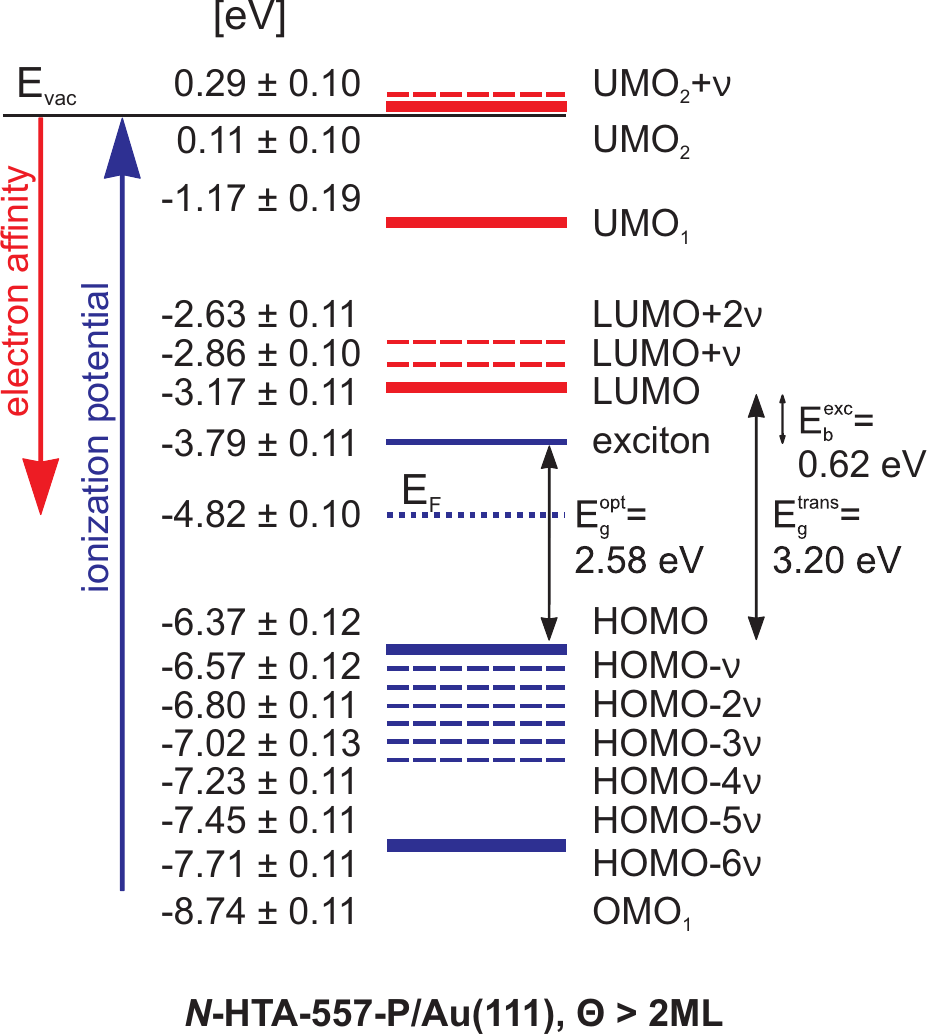}}
\caption{Energy level diagram of  N-HTA-557-P adsorbed on Au(111) for N-HTA-557-P coverages above 2 ML with respect to the vacuum level. The blue levels are the ionization potentials and the red ones are the electron affinities, which are in the molecular orbital picture represented by the energies of the HOMO, LUMO, UMOs (unoccupied molecular orbital) or OMO (occupied molecular orbital), respectively. The dashed lines indicate vibronic transition. $E_{F}$ is the Au(111) Fermi level.}
\label{energydiagram}
\end{figure}

In Figure \ref{energydiagram}, we summarize the level alignments with respect to the vacuum level. Note that the work function of bare Au(111) ($\Phi_{Au(111)}$ = 5.5 eV) decreases by 0.7 eV due to the adsorption of 1 ML N-HTA-557-P (see supporting information, Figs. S1 and S2) and stays constant for higher coverages. As discussed before \cite{Varene2011, Bogner2016, Bogner2015, Bronner2016, Stein2019}, depending on the excitation mechanism 2PPE probes transport levels.
Unoccupied single-electron states can be populated via an electron transfer from the metal to the N-HTA-557-P molecule creating a negative ion resonance and occupied  single-electron states are ionized, creating a positive ion resonance.
Thus, we obtain the electron affinities (EA) and the ionization potentials (IP) of the investigated  molecule.
In addition, in 2PPE also intramolecular electronic excitation is possible, which corresponds to exciton generation, i.e. formation of an electron-hole pair, in which the molecule remains overall neutral.
The minimal energy required for this process is the optical gap ($E_{opt}$), which corresponds to the excitation energy of the lowest excited singlet state.
$E_{opt}$ is smaller than the difference between the lowest IP
and the highest EA.
The so-called transport gap is given as $E_{transp} = IP - EA$.
The difference between the transport gap and the optical gap corresponds to the exciton binding energy $E_{transp}-E_{opt}=E_{B}$ \cite{Muntwiler2010, Knupfer2003}.
For the N-HTA-557-P  we obtain a transport gap of 3.20 eV (IP = 6.37 eV, EA = 3.17 eV), and an optical gap of 2.58 eV, the excitation energy of the lowest excited singlet state.
Hence, the exciton binding energy ($E_{B}$) of this lowest excitonic state amounts to 620 meV. For comparison, N-HTA-557-P in solution (CH$_2$Cl$_2$) possesses an optical gap of 2.61 eV (see supporting information, Fig. S3). The similar value indicates that the adsorbed molecules (5 ML) are decoupled from the metallic substrate and weak adsorbate-adsorbate (lateral) interactions.
In addition, we identified two further unoccupied molecular states (unoccupied molecular orbital: UMO), an intermediate (at 1.17 eV)  and a final state (0.11 eV above $E_{vac}$) as well as an occupied molecular state (occupied molecular orbital: OMO) at 8.74 eV. All other 2PPE features are assigned to vibronic contributions related to the LUMO, and UMO$_{2}$ as well as to the HOMO. They exhibit an equidistance energy different of around 200 meV, which can be related to an excitation of $\nu$(C--C) and $\delta$(C--H) vibrational modes of the N-HTA-557 moiety of N-HTA-557-P located at 1636 cm$^{-1}$  (see supporting information Fig. S4 and Tab. S1). Vibronic features in 2PPE data have recently been also proposed for another N-heterocyclic compound adsorbed on Au(111) supported by scanning tunneling spectroscopy results \cite{Stein2021b}. For lower N-HTA-557-P coverages, i.e., around 1ML conributions from the Au(111) surface, the \emph{d}-bands, the shifted surface state and the first image potential state are visible in the 2PPE data (see supporting information, Fig. S1).

 \begin{figure}[h!]
\centering
\resizebox{0.75\hsize}{!}{\includegraphics[width=1\textwidth]{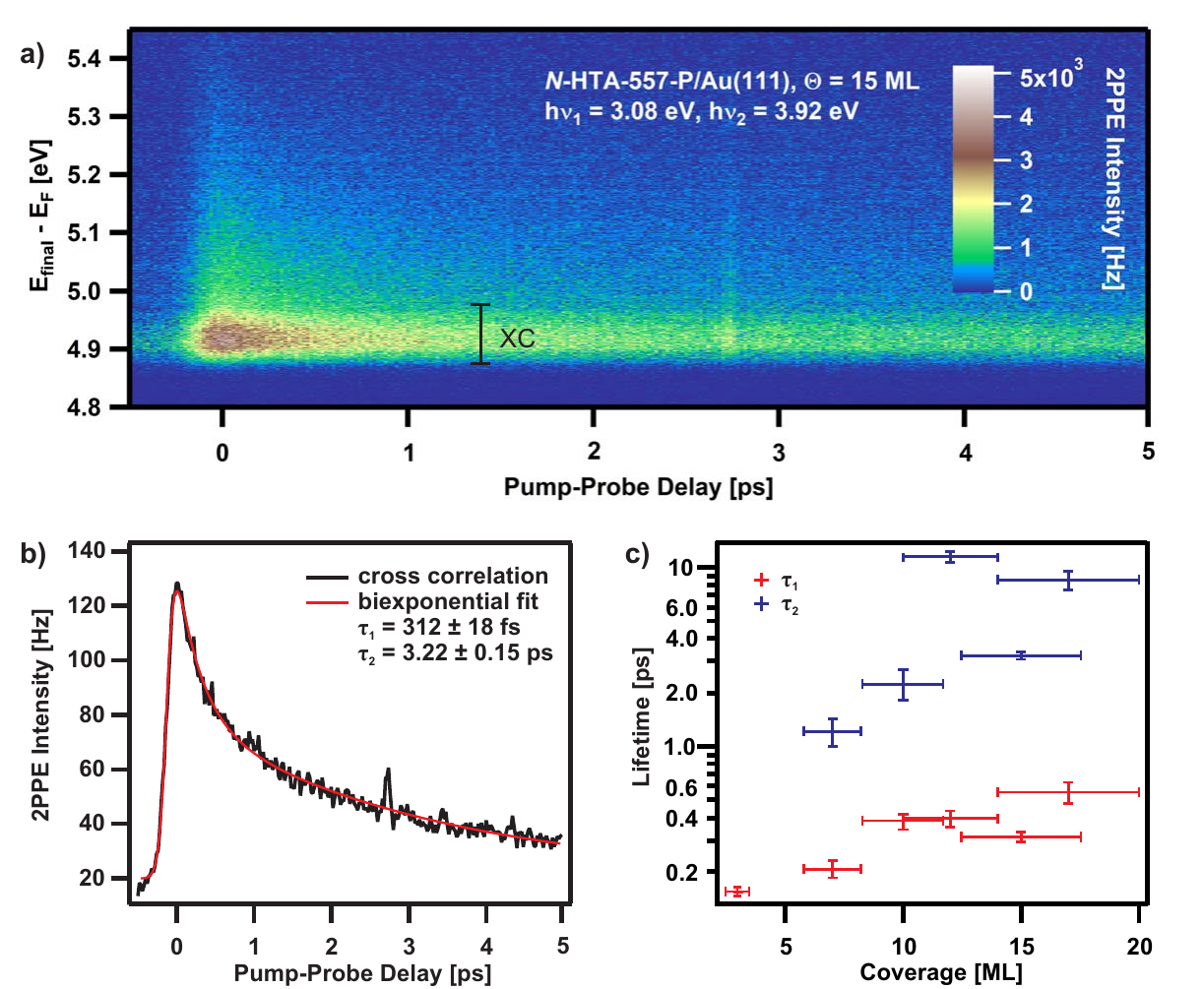}}
\caption{a) 2D-plot of the 2PPE intensity as a function of final-state energy and the pump-probe delay for 15 ML N-HTA-557-P on Au(111). b)The cross-correlation (XC) curves fitted by a $sech^{2}$ function convoluted by a double exponential decay (XC energy
range in a)). c) Coverage dependent lifetimes.}
\label{TR-2PPE}
\end{figure}
We used femtosecond time-resolved 2PPE to study how the electronic coupling
between the N-HTA-557-P molecules and the metal substrate influences the dynamics of
optically excited states such as excitons in N-HTA-557-P. For low coverages (1--2 ML) none of the electronically excited states, e.g. the LUMO or excitonic states, exhibit a detectable lifetime, thus $\tau$ $<$ 10 fs. The reason for these ultrashort lifetimes is a strong electron coupling enabling an efficient electron transfer from the molecule back to the metal. Such short lifetime in the order of a few femtoseconds have been found also for other adsorbates on metal surfaces \cite{Yang2008, Dutton2005, Hagen2010, Bronner2012a}. In contrast, for higher coverages longer lifetimes are usually observed. This is the case for our findings as can be seen Fig. \ref{TR-2PPE}a for 15 ML N-HTA-557-P on Au(111)
 in a two-dimensional false colored plot representation taken at a pump photon energy of $h\nu_1$ = 3.08 eV and a probe photon energy $h\nu_2$ = 3.92 eV. Such representations visualize the normalized correlated dichromatic 2PPE signal at a given final state energy ($E_{Final}$) with respect to $E_{F}$ as a function of pump-probe delay. Positive pump-probe delays imply that the pump pulse h$\nu_1$ reaches the sample after the probe pulse h$\nu_2$, and \emph{vice versa} for negative delays. A long-lived contribution can clearly be seen in the energy region around 4.9 eV, thus the energy at which we identified the first excitonic state. To receive the lifetimes of the involved state a cross-correlation (XC) curve as shown in Fig. \ref{TR-2PPE}b has been employed. To fit the XC curve, we used a $sech^{2}$-function representing the laser pulse duration convoluted with a response function of the
intermediate state. A superposition of two exponential decays with different time constants, $\tau_{1}$ and $\tau_{2}$, describes the time-resolved
photoemission data well (see solid line in Fig. \ref{TR-2PPE} b)). The time constants
 are $\tau_1$ = 312$\pm$18 fs and $\tau_2$ = 3.22$\pm$0.15 ps at a coverage of 15 ML. Coverage-dependent measurements (see Fig.\ref{TR-2PPE}c) clearly demonstrate  that the lifetimes increase with rising coverage.
 Coverage-dependent lifetimes can be explained by the availability of two relaxation channels \cite{Dutton2005, Varene2012, Bogner2015, Bogner2016, Stein2017}. An intrinsic channel due to the decay in the bulk material and a distance-dependent (external) channel resulting from a quenching process by the metallic substrate, i.e, transfer of the electrons to the metal. Note that, in literature no time-dependent measurements on the N-HTAs excited states dynamics are available. However, by comparing with our previous time-resolved 2PPE results obtained from other electron donating materials \cite{Bogner2015, Bogner2016} we consider the following processes: Excitation with a photon energy of 3.08 eV leads to the creation of hot ($E_{opt.} = 2.58 eV$), delocalized excitons. These excitons relax and localize on an ultrafast time-scale ($\tau_1$ = 312$\pm$18 fs). A very fast relaxation of hot excitons has also been found in other organic semiconductors \cite{Marks2013, Engel2005, Guo2009, Chen2013}. The longer lived component ($\tau_2$ = 3.22$\pm$0.15 ps) may be related to the decay of the exciton to the ground state. However, it could also be connected to the lifetime of polarons as well as electrons bound at defect sites. Note, that no energetic stabilization is observed, which would be expected for polaron formation or population of defect sites.

In summary, we employed energy- and time-resolved two-photon photoemission spectroscopy to quantitatively determine the energetic positions of unoccupied and occupied electronic states as well as an excitonic state and to resolve the exciton dynamics in a novel $\pi$-expanded N-heterotriangulene derivative adsorbed on Au(111). A transport gap of 3.20 eV and an optical gap of 2.58 eV have been identified, resulting in an exciton binding energy of 0.62 eV. Hot and delocalized excitons have found to localize on a femtosecond timescale followed by an exciton decay within picoseconds.
The quantitative determination of the energetic
positions of molecular electronic levels, the achieved
detailed understanding of the electronic structure and insights into the exciton dynamics is an
essential prerequisite for the design of new functional organic
molecules for (opto)electronic devices.

\section{Experimental Methods}
2PPE measurements were conducted in an ultra-high vacuum chamber. A clean Au(111) substrate was prepared by a standard procedure including Ar$^+$ sputtering and subsequent annealing to 800 K under ultra-high vacuum
conditions. The N-HTA-557-P molecules were synthesized according to Ref. \cite{Michalsky2022}, purified by column chromatography and degassed in vacuum at 400 K for several hours. They were deposited from an effusion cell held at a temperature of 453 K while the surface was kept at 100 K. The experiments were performed at a sample temperature of around 90 K. In 2PPE, a femtosecond pump laser pulse induces electronic transitions from occupied electronic states of the substrate or the adsorbate into unoccupied
electronic or virtual states. These excited states are then probed with a second laser pulse
by lifting the excited electron above the vacuum level.
The kinetic energy of the emitted electrons is measured with
a time-of-flight spectrometer. Photoelectron signals may arise from the
occupied or unoccupied electronic states; therefore, photon
energy-dependent measurements are needed for the assignment
(for details, see refs see \cite{Bogner2016, Bogner2015, Tegeder2012}). A temporal delay between pump
and probe pulse was introduced to obtain population dynamics
of excited states on a femtosecond time scale \cite{Varene2012, Bogner2016}.

\begin{acknowledgement}
Funding by the German Research Foundation (DFG) through
the collaborative research center SFB 1249 "N-Heteropolycycles as Functional Materials" (Project Number 281029004-SFB 1249, projects A05 and B06) is gratefully acknowledged.
\end{acknowledgement}

\begin{suppinfo}
Two-photon photoemission data, UV/vis absorption and emission data as well as infrared vibrational data.
\end{suppinfo}

\bibliography{N-HTA_lit}

\providecommand{\latin}[1]{#1}
\makeatletter
\providecommand{\doi}
  {\begingroup\let\do\@makeother\dospecials
  \catcode`\{=1 \catcode`\}=2 \doi@aux}
\providecommand{\doi@aux}[1]{\endgroup\texttt{#1}}
\makeatother
\providecommand*\mcitethebibliography{\thebibliography}
\csname @ifundefined\endcsname{endmcitethebibliography}
  {\let\endmcitethebibliography\endthebibliography}{}
\begin{mcitethebibliography}{48}
\providecommand*\natexlab[1]{#1}
\providecommand*\mciteSetBstSublistMode[1]{}
\providecommand*\mciteSetBstMaxWidthForm[2]{}
\providecommand*\mciteBstWouldAddEndPuncttrue
  {\def\EndOfBibitem{\unskip.}}
\providecommand*\mciteBstWouldAddEndPunctfalse
  {\let\EndOfBibitem\relax}
\providecommand*\mciteSetBstMidEndSepPunct[3]{}
\providecommand*\mciteSetBstSublistLabelBeginEnd[3]{}
\providecommand*\EndOfBibitem{}
\mciteSetBstSublistMode{f}
\mciteSetBstMaxWidthForm{subitem}{(\alph{mcitesubitemcount})}
\mciteSetBstSublistLabelBeginEnd
  {\mcitemaxwidthsubitemform\space}
  {\relax}
  {\relax}

\bibitem[Klauk \latin{et~al.}(2007)Klauk, Zschieschang, Pflaum, and
  Halik]{klauk2007}
Klauk,~H.; Zschieschang,~U.; Pflaum,~J.; Halik,~M. Ultralow-Power Organic
  Complementary Circuits. \emph{Nature} \textbf{2007}, \emph{445},
  745--748\relax
\mciteBstWouldAddEndPuncttrue
\mciteSetBstMidEndSepPunct{\mcitedefaultmidpunct}
{\mcitedefaultendpunct}{\mcitedefaultseppunct}\relax
\EndOfBibitem
\bibitem[W{\"u}rthner and Stolte(2011)W{\"u}rthner, and Stolte]{wurthner2011}
W{\"u}rthner,~F.; Stolte,~M. Naphthalene and Perylene Diimides for Organic
  Transistors. \emph{Chem. Commun.} \textbf{2011}, \emph{47}, 5109--5115\relax
\mciteBstWouldAddEndPuncttrue
\mciteSetBstMidEndSepPunct{\mcitedefaultmidpunct}
{\mcitedefaultendpunct}{\mcitedefaultseppunct}\relax
\EndOfBibitem
\bibitem[Bunz \latin{et~al.}(2013)Bunz, Engelhart, Lindner, and
  Schaffroth]{Bunz2013}
Bunz,~U. H.~F.; Engelhart,~J.~U.; Lindner,~B.~D.; Schaffroth,~M. Large
  N-Heteroacenes: New Tricks for Very Old Dogs? \emph{Angew. Chem. Int. Ed.}
  \textbf{2013}, \emph{52}, 3810--3821\relax
\mciteBstWouldAddEndPuncttrue
\mciteSetBstMidEndSepPunct{\mcitedefaultmidpunct}
{\mcitedefaultendpunct}{\mcitedefaultseppunct}\relax
\EndOfBibitem
\bibitem[Miao(2014)]{miao2014}
Miao,~Q. Ten Years of N-Heteropentacenes as Semiconductors for Organic
  Thin-Film Transistors. \emph{Adv. Mater.} \textbf{2014}, \emph{26},
  5541--5549\relax
\mciteBstWouldAddEndPuncttrue
\mciteSetBstMidEndSepPunct{\mcitedefaultmidpunct}
{\mcitedefaultendpunct}{\mcitedefaultseppunct}\relax
\EndOfBibitem
\bibitem[Bunz(2015)]{bunz2015}
Bunz,~U. H.~F. The Larger Linear N-Heteroacenes. \emph{Acc. Chem. Res.}
  \textbf{2015}, \emph{48}, 1676--1686\relax
\mciteBstWouldAddEndPuncttrue
\mciteSetBstMidEndSepPunct{\mcitedefaultmidpunct}
{\mcitedefaultendpunct}{\mcitedefaultseppunct}\relax
\EndOfBibitem
\bibitem[Hammer \latin{et~al.}(2015)Hammer, Schaub, Meinhardt, and
  Kivala]{Hammer2015}
Hammer,~N.; Schaub,~T.~A.; Meinhardt,~U.; Kivala,~M. N-Heterotriangulenes:
  Fascinating Relatives of Triphenylamine. \emph{Chem. Rec.} \textbf{2015},
  \emph{15}, 1119--1131\relax
\mciteBstWouldAddEndPuncttrue
\mciteSetBstMidEndSepPunct{\mcitedefaultmidpunct}
{\mcitedefaultendpunct}{\mcitedefaultseppunct}\relax
\EndOfBibitem
\bibitem[Michalsky \latin{et~al.}(2022)Michalsky, Gensch, Walla, Hoffmann,
  Rominger, Oeser, Tegeder, Dreuw, and Kivala]{Michalsky2022}
Michalsky,~I.; Gensch,~V.; Walla,~C.; Hoffmann,~M.; Rominger,~F.; Oeser,~T.;
  Tegeder,~P.; Dreuw,~A.; Kivala,~M. Fully Bridged Triphenylamines Comprising
  Five- and Seven-Membered Rings. \emph{Chem. Eur. J.} \textbf{2022},
  \emph{28}, e202200326\relax
\mciteBstWouldAddEndPuncttrue
\mciteSetBstMidEndSepPunct{\mcitedefaultmidpunct}
{\mcitedefaultendpunct}{\mcitedefaultseppunct}\relax
\EndOfBibitem
\bibitem[Wagner \latin{et~al.}(2022)Wagner, Crocomo, Kochman, Kubas, Data, and
  Lindner]{Wagner2022}
Wagner,~J.; Crocomo,~P.~Z.; Kochman,~M.~A.; Kubas,~A.; Data,~P.; Lindner,~M.
  Modular Nitrogen-Doped Concave Polycyclic Aromatic Hydrocarbons for
  High-Performance Organic Light-Emitting Diodes with Tunable Emission
  Mechanisms. \emph{Angew. Chem. Int. Ed.} \textbf{2022}, \emph{61},
  e202202232\relax
\mciteBstWouldAddEndPuncttrue
\mciteSetBstMidEndSepPunct{\mcitedefaultmidpunct}
{\mcitedefaultendpunct}{\mcitedefaultseppunct}\relax
\EndOfBibitem
\bibitem[Meinhardt \latin{et~al.}(2016)Meinhardt, Lodermeyer, Schaub, Kunzmann,
  Dral, Sale, Hampel, Guldi, Costa, and Kivala]{Meinhardt2016}
Meinhardt,~U.; Lodermeyer,~F.; Schaub,~T.~A.; Kunzmann,~A.; Dral,~P.~O.;
  Sale,~A.~C.; Hampel,~F.; Guldi,~D.~M.; Costa,~R.~D.; Kivala,~M.
  N-Heterotriangulene Chromophores with 4-Pyridyl Anchors for Dye-Sensitized
  Solar Cells. \emph{RSC Adv.} \textbf{2016}, \emph{6}, 67372--67377\relax
\mciteBstWouldAddEndPuncttrue
\mciteSetBstMidEndSepPunct{\mcitedefaultmidpunct}
{\mcitedefaultendpunct}{\mcitedefaultseppunct}\relax
\EndOfBibitem
\bibitem[Hirai \latin{et~al.}(2019)Hirai, Tanaka, Sakai, and
  Yamaguchi]{Hirai2019}
Hirai,~M.; Tanaka,~N.; Sakai,~M.; Yamaguchi,~S. Structurally Constrained
  Boron-, Nitrogen-, Silicon-, and Phosphorus-Centered Polycyclic
  $\pi$-Conjugated Systems. \emph{Chem. Rev.} \textbf{2019}, \emph{119},
  8291--8331\relax
\mciteBstWouldAddEndPuncttrue
\mciteSetBstMidEndSepPunct{\mcitedefaultmidpunct}
{\mcitedefaultendpunct}{\mcitedefaultseppunct}\relax
\EndOfBibitem
\bibitem[Schaub \latin{et~al.}(2020)Schaub, Padberg, and Kivala]{Schaub2020}
Schaub,~T.~A.; Padberg,~K.; Kivala,~M. Bridged Triarylboranes, -Silanes,
  -Amines, and -Phosphines as Minimalistic Heteroatom-Containing Polycyclic
  Aromatic Hydrocarbons: Progress and Challenges. \emph{J. Phys. Org. Chem.}
  \textbf{2020}, \emph{33}, e4022\relax
\mciteBstWouldAddEndPuncttrue
\mciteSetBstMidEndSepPunct{\mcitedefaultmidpunct}
{\mcitedefaultendpunct}{\mcitedefaultseppunct}\relax
\EndOfBibitem
\bibitem[Krug \latin{et~al.}(2020)Krug, Wagner, Schaub, Zhang,
  Sch\"{u}{\ss}lbauer, Ascherl, M\"{u}nich, Schr\"{o}der, Gr\"{o}hn, Dral, and
  et~al.]{Krug2020}
Krug,~M.; Wagner,~M.; Schaub,~T.~A.; Zhang,~W.; Sch\"{u}{\ss}lbauer,~C.~M.;
  Ascherl,~J. D.~R.; M\"{u}nich,~P.~W.; Schr\"{o}der,~R.~R.; Gr\"{o}hn,~F.;
  Dral,~P.~O.; et~al., Quantification of a CH–$\pi$ Interaction Responsible
  for Chiral Discrimination and Evaluation of Its Contribution to
  Enantioselectivity. \emph{Angew. Chem. Int. Ed.} \textbf{2020}, \emph{59},
  16233--16240\relax
\mciteBstWouldAddEndPuncttrue
\mciteSetBstMidEndSepPunct{\mcitedefaultmidpunct}
{\mcitedefaultendpunct}{\mcitedefaultseppunct}\relax
\EndOfBibitem
\bibitem[Kader \latin{et~al.}(2019)Kader, St\"{o}ger, Fr\"{o}hlich, and
  Kautny]{Kader2019}
Kader,~T.; St\"{o}ger,~B.; Fr\"{o}hlich,~J.; Kautny,~P.
  Azaindolo[3,2,1-jk]carbazoles: New Building Blocks for Functional Organic
  Materials. \emph{Chem. Eur. J.} \textbf{2019}, \emph{25}, 4412--4425\relax
\mciteBstWouldAddEndPuncttrue
\mciteSetBstMidEndSepPunct{\mcitedefaultmidpunct}
{\mcitedefaultendpunct}{\mcitedefaultseppunct}\relax
\EndOfBibitem
\bibitem[Ajdari \latin{et~al.}(2023)Ajdari, Pappenberger, Walla, Hoffmann,
  Michalsky, Kivala, Dreuw, and Tegeder]{Ajdari2021b}
Ajdari,~M.; Pappenberger,~R.; Walla,~C.; Hoffmann,~M.; Michalsky,~I.;
  Kivala,~M.; Dreuw,~A.; Tegeder,~P. The Impact of Connectivity on the
  Electronic Structure of N-Heterotriangulenes. \emph{J. Phys. Chem. C}
  \textbf{2023}, \emph{127}, 542--549\relax
\mciteBstWouldAddEndPuncttrue
\mciteSetBstMidEndSepPunct{\mcitedefaultmidpunct}
{\mcitedefaultendpunct}{\mcitedefaultseppunct}\relax
\EndOfBibitem
\bibitem[Deng and Zhang(2019)Deng, and Zhang]{Deng2019}
Deng,~N.; Zhang,~G. Nitrogen-Centered Concave Molecules with Double Fused
  Pentagons. \emph{Org. Lett.} \textbf{2019}, \emph{21}, 5248--5251\relax
\mciteBstWouldAddEndPuncttrue
\mciteSetBstMidEndSepPunct{\mcitedefaultmidpunct}
{\mcitedefaultendpunct}{\mcitedefaultseppunct}\relax
\EndOfBibitem
\bibitem[Zhu(2004)]{Zhu2004}
Zhu,~X.-Y. Electronic Structure and Electron Dynamics at Molecule-Metal
  Interfaces: Implications for Molecule-Based Electronics. \emph{Surf. Sci.
  Rep.} \textbf{2004}, \emph{56}, 1--83\relax
\mciteBstWouldAddEndPuncttrue
\mciteSetBstMidEndSepPunct{\mcitedefaultmidpunct}
{\mcitedefaultendpunct}{\mcitedefaultseppunct}\relax
\EndOfBibitem
\bibitem[Zhu(2004)]{Zhu2004a}
Zhu,~X.-Y. Charge Transport at Metal-Molecule Interfaces: {A} Spectroscopic
  View. \emph{J. Phys. Chem. B} \textbf{2004}, \emph{108}, 8778--8793\relax
\mciteBstWouldAddEndPuncttrue
\mciteSetBstMidEndSepPunct{\mcitedefaultmidpunct}
{\mcitedefaultendpunct}{\mcitedefaultseppunct}\relax
\EndOfBibitem
\bibitem[Lindstrom and Zhu(2006)Lindstrom, and Zhu]{Lindstrom2006}
Lindstrom,~C.; Zhu,~X.-Y. Photoinduced Electron Transfer at Molecule-Metal
  Interfaces. \emph{Chem. Rev.} \textbf{2006}, \emph{106}, 4281--4300\relax
\mciteBstWouldAddEndPuncttrue
\mciteSetBstMidEndSepPunct{\mcitedefaultmidpunct}
{\mcitedefaultendpunct}{\mcitedefaultseppunct}\relax
\EndOfBibitem
\bibitem[Muntwiler and Zhu(2010)Muntwiler, and Zhu]{Muntwiler2010}
Muntwiler,~M.; Zhu,~X.-Y. In \emph{Dynamics at Solid State Surfaces and
  Interfaces}; Bovensiepen,~U., Petek,~H., Wolf,~M., Eds.; Wiley-VCH,
  2010\relax
\mciteBstWouldAddEndPuncttrue
\mciteSetBstMidEndSepPunct{\mcitedefaultmidpunct}
{\mcitedefaultendpunct}{\mcitedefaultseppunct}\relax
\EndOfBibitem
\bibitem[Bronner \latin{et~al.}(2012)Bronner, Schulze, Hagen, and
  Tegeder]{bronner2012influence}
Bronner,~C.; Schulze,~M.; Hagen,~S.; Tegeder,~P. The Influence of the
  Electronic Structure of Adsorbate--Substrate Complexes on Photoisomerization
  Ability. \emph{New J. Phys.} \textbf{2012}, \emph{14}, 043023\relax
\mciteBstWouldAddEndPuncttrue
\mciteSetBstMidEndSepPunct{\mcitedefaultmidpunct}
{\mcitedefaultendpunct}{\mcitedefaultseppunct}\relax
\EndOfBibitem
\bibitem[Bronner \latin{et~al.}(2011)Bronner, Schulze, Franke, Pascual, and
  Tegeder]{bronner2011switching}
Bronner,~C.; Schulze,~G.; Franke,~K.~J.; Pascual,~J.~I.; Tegeder,~P. Switching
  Ability of Nitro-Spiropyran on {Au(111): Electronic} Structure Changes as a
  Sensitive Probe During a Ring-Opening Reaction. \emph{J. Phys.: Condens.
  Matter} \textbf{2011}, \emph{23}, 484005\relax
\mciteBstWouldAddEndPuncttrue
\mciteSetBstMidEndSepPunct{\mcitedefaultmidpunct}
{\mcitedefaultendpunct}{\mcitedefaultseppunct}\relax
\EndOfBibitem
\bibitem[Armbrust \latin{et~al.}(2012)Armbrust, G\"{u}dde, Jakob, and
  H\"{o}fer]{Armbrust2012}
Armbrust,~N.; G\"{u}dde,~J.; Jakob,~P.; H\"{o}fer,~U. Time-Resolved Two-Photon
  Photoemission of Unoccupied Electronic States of Periodically Rippled
  Graphene on Ru(0001). \emph{Phys. Rev. Lett.} \textbf{2012}, \emph{108},
  056801\relax
\mciteBstWouldAddEndPuncttrue
\mciteSetBstMidEndSepPunct{\mcitedefaultmidpunct}
{\mcitedefaultendpunct}{\mcitedefaultseppunct}\relax
\EndOfBibitem
\bibitem[Bronner \latin{et~al.}(2013)Bronner, Stremlau, Gille, Brau{\ss}e,
  Haase, Hecht, and Tegeder]{bronner2013}
Bronner,~C.; Stremlau,~S.; Gille,~M.; Brau{\ss}e,~F.; Haase,~A.; Hecht,~S.;
  Tegeder,~P. Aligning the Band Gap of Graphene Nanoribbons by Monomer Doping.
  \emph{Angew. Chem. Int. Ed.} \textbf{2013}, \emph{52}, 4422--4425\relax
\mciteBstWouldAddEndPuncttrue
\mciteSetBstMidEndSepPunct{\mcitedefaultmidpunct}
{\mcitedefaultendpunct}{\mcitedefaultseppunct}\relax
\EndOfBibitem
\bibitem[Bogner \latin{et~al.}(2016)Bogner, Yang, Baum, M.~Corso, B\"{a}uerle,
  Franke, Pascual, and Tegeder]{Bogner2016}
Bogner,~L.; Yang,~Z.; Baum,~S.; M.~Corso,~R.~F.; B\"{a}uerle,~P.;
  Franke,~K.~J.; Pascual,~J.~I.; Tegeder,~P. Electronic States and Exciton
  Dynamics in Dicyanovinyl-Sexithiophene on {Au(111)}. \emph{J. Phys. Chem. C}
  \textbf{2016}, \emph{120}, 27268--27275\relax
\mciteBstWouldAddEndPuncttrue
\mciteSetBstMidEndSepPunct{\mcitedefaultmidpunct}
{\mcitedefaultendpunct}{\mcitedefaultseppunct}\relax
\EndOfBibitem
\bibitem[Bogner \latin{et~al.}(2015)Bogner, Yang, Corso, Fitzner, B{\"a}uerle,
  Franke, Pascual, and Tegeder]{Bogner2015}
Bogner,~L.; Yang,~Z.; Corso,~M.; Fitzner,~R.; B{\"a}uerle,~P.; Franke,~K.~J.;
  Pascual,~J.~I.; Tegeder,~P. Electronic Structure and Excited States Dynamics
  in a Dicyanovinyl-Substituted Oligothiophene on {Au(111)}. \emph{Phys. Chem.
  Chem. Phys.} \textbf{2015}, \emph{17}, 27118--27126\relax
\mciteBstWouldAddEndPuncttrue
\mciteSetBstMidEndSepPunct{\mcitedefaultmidpunct}
{\mcitedefaultendpunct}{\mcitedefaultseppunct}\relax
\EndOfBibitem
\bibitem[Stein \latin{et~al.}(2017)Stein, Maass, and Tegeder]{Stein2017}
Stein,~A.; Maass,~F.; Tegeder,~P. Triisopropylsilylethynyl-Pentacene on
  Au(111): Adsorption Properties, Electronic Structure and Singlet-Fission
  Dynamics. \emph{J. Phys. Chem. C} \textbf{2017}, \emph{121},
  18075--18083\relax
\mciteBstWouldAddEndPuncttrue
\mciteSetBstMidEndSepPunct{\mcitedefaultmidpunct}
{\mcitedefaultendpunct}{\mcitedefaultseppunct}\relax
\EndOfBibitem
\bibitem[Gerbert \latin{et~al.}(2017)Gerbert, Maass, and Tegeder]{Gerbert2017}
Gerbert,~D.; Maass,~F.; Tegeder,~P. Extended Space Charge Region and Unoccupied
  Molecular Band Formation in Epitaxial Tetrafluoro-tetracyanoquinodimethane
  Films. \emph{J. Phys. Chem. C} \textbf{2017}, \emph{121}, 15696--15701\relax
\mciteBstWouldAddEndPuncttrue
\mciteSetBstMidEndSepPunct{\mcitedefaultmidpunct}
{\mcitedefaultendpunct}{\mcitedefaultseppunct}\relax
\EndOfBibitem
\bibitem[Gerbert and Tegeder(2017)Gerbert, and Tegeder]{Gerbert2017a}
Gerbert,~D.; Tegeder,~P. Molecular Ion Formation by Photo-Induced Electron
  Transfer at the Tetracyanoquinodimethane/Au(111) Interface. \emph{J. Phys.
  Chem. Lett.} \textbf{2017}, \emph{8}, 4685--4690\relax
\mciteBstWouldAddEndPuncttrue
\mciteSetBstMidEndSepPunct{\mcitedefaultmidpunct}
{\mcitedefaultendpunct}{\mcitedefaultseppunct}\relax
\EndOfBibitem
\bibitem[Bronsch \latin{et~al.}(2019)Bronsch, Wagner, Baum, Wansleben, Zielke,
  Ghanbari, Gy\"{o}r\"{o}k, Navarro-Quezada, Zeppenfeld, Weinelt, and
  et~al.]{Bronsch2019}
Bronsch,~W.; Wagner,~T.; Baum,~S.; Wansleben,~M.; Zielke,~K.; Ghanbari,~E.;
  Gy\"{o}r\"{o}k,~M.; Navarro-Quezada,~A.; Zeppenfeld,~P.; Weinelt,~M.; et~al.,
  Interplay between Morphology and Electronic Structure in Sexithiophene Films
  on Au(111). \emph{J. Phys. Chem. C} \textbf{2019}, \emph{123},
  7931--7939\relax
\mciteBstWouldAddEndPuncttrue
\mciteSetBstMidEndSepPunct{\mcitedefaultmidpunct}
{\mcitedefaultendpunct}{\mcitedefaultseppunct}\relax
\EndOfBibitem
\bibitem[Stein \latin{et~al.}(2019)Stein, Rolf, Lotze, Czekelius, Franke, and
  Tegeder]{Stein2019}
Stein,~A.; Rolf,~D.; Lotze,~C.; Czekelius,~C.; Franke,~K.~J.; Tegeder,~P.
  Electronic Structure of an Iron-Porphyrin Derivative on Au(111). \emph{J.
  Phys.: Condens Matter} \textbf{2019}, \emph{31}, 044002\relax
\mciteBstWouldAddEndPuncttrue
\mciteSetBstMidEndSepPunct{\mcitedefaultmidpunct}
{\mcitedefaultendpunct}{\mcitedefaultseppunct}\relax
\EndOfBibitem
\bibitem[Ajdari \latin{et~al.}(2020)Ajdari, Stein, Hoffmann, M\"{u}ller, Bunz,
  Dreuw, and Tegeder]{Ajdari2020}
Ajdari,~M.; Stein,~A.; Hoffmann,~M.; M\"{u}ller,~M.; Bunz,~U. H.~F.; Dreuw,~A.;
  Tegeder,~P. Lightening up a Dark State of a Pentacene Derivative via
  N-Introduction. \emph{J. Phys. Chem. C} \textbf{2020}, \emph{124},
  7196--7204\relax
\mciteBstWouldAddEndPuncttrue
\mciteSetBstMidEndSepPunct{\mcitedefaultmidpunct}
{\mcitedefaultendpunct}{\mcitedefaultseppunct}\relax
\EndOfBibitem
\bibitem[Stein \latin{et~al.}(2021)Stein, Rolf, Lotze, G\"{u}nther, Gade,
  Franke, and Tegeder]{Stein2021}
Stein,~A.; Rolf,~D.; Lotze,~C.; G\"{u}nther,~B.; Gade,~L.~H.; Franke,~K.~J.;
  Tegeder,~P. Band Formation at Interfaes Between N-Heteropolycycles and Gold
  Eletrodes. \emph{J. Phys. Chem. Lett.} \textbf{2021}, \emph{12},
  947--951\relax
\mciteBstWouldAddEndPuncttrue
\mciteSetBstMidEndSepPunct{\mcitedefaultmidpunct}
{\mcitedefaultendpunct}{\mcitedefaultseppunct}\relax
\EndOfBibitem
\bibitem[Stein \latin{et~al.}(2021)Stein, Rolf, Lotze, Feldmann, Gerbert,
  G\"{u}nther, Jeindl, Cartus, Hofmann, Gade, and et~al.]{Stein2021b}
Stein,~A.; Rolf,~D.; Lotze,~C.; Feldmann,~S.; Gerbert,~D.; G\"{u}nther,~B.;
  Jeindl,~A.; Cartus,~J.~J.; Hofmann,~O.~T.; Gade,~L.~H.; et~al., Electronic
  Properties of Tetraazaperopyrene Derivatives on Au(111): Energy Level
  Alignment and Interfacial Band Formation. \emph{J. Phys. Chem. C}
  \textbf{2021}, \emph{125}, 19969--19979\relax
\mciteBstWouldAddEndPuncttrue
\mciteSetBstMidEndSepPunct{\mcitedefaultmidpunct}
{\mcitedefaultendpunct}{\mcitedefaultseppunct}\relax
\EndOfBibitem
\bibitem[Stallberg \latin{et~al.}(2022)Stallberg, Namgalies, Chatterjee, and
  H\"{o}fer]{Stallberg2022}
Stallberg,~K.; Namgalies,~A.; Chatterjee,~S.; H\"{o}fer,~U. Ultrafast Exciton
  Dynamics and Charge Transfer at PTCDA/Metal Interfaces. \emph{J. Phys. Chem.
  C} \textbf{2022}, \emph{126}, 12728--12734\relax
\mciteBstWouldAddEndPuncttrue
\mciteSetBstMidEndSepPunct{\mcitedefaultmidpunct}
{\mcitedefaultendpunct}{\mcitedefaultseppunct}\relax
\EndOfBibitem
\bibitem[Bronner \latin{et~al.}(2012)Bronner, Schulze, Hagen, and
  Tegeder]{Bronner2012a}
Bronner,~C.; Schulze,~M.; Hagen,~S.; Tegeder,~P. The Influence of the
  Electronic Structure of Adsorbate-Substrate Complexes on the
  Photoisomerization Ability. \emph{New J. Phys.} \textbf{2012}, \emph{14},
  043032\relax
\mciteBstWouldAddEndPuncttrue
\mciteSetBstMidEndSepPunct{\mcitedefaultmidpunct}
{\mcitedefaultendpunct}{\mcitedefaultseppunct}\relax
\EndOfBibitem
\bibitem[Varene \latin{et~al.}(2011)Varene, Martin, and Tegeder]{Varene2011}
Varene,~E.; Martin,~I.; Tegeder,~P. Optically Induced Inter- and Intrafacial
  Electron Transfer Probed by Two-Photon Photoemission: Electronic States of
  Sexithiophene on {Au(111)}. \emph{J. Phys. Chem. Lett.} \textbf{2011},
  \emph{2}, 252--256\relax
\mciteBstWouldAddEndPuncttrue
\mciteSetBstMidEndSepPunct{\mcitedefaultmidpunct}
{\mcitedefaultendpunct}{\mcitedefaultseppunct}\relax
\EndOfBibitem
\bibitem[Bronner \latin{et~al.}(2016)Bronner, Gerbert, Broska, and
  Tegeder]{Bronner2016}
Bronner,~C.; Gerbert,~D.; Broska,~A.; Tegeder,~P. Excitonic States in Narrow
  Armchair Graphene Nanoribbons on Gold Surfaces. \emph{J. Phys. Chem. C}
  \textbf{2016}, \emph{120}, 26168--26172\relax
\mciteBstWouldAddEndPuncttrue
\mciteSetBstMidEndSepPunct{\mcitedefaultmidpunct}
{\mcitedefaultendpunct}{\mcitedefaultseppunct}\relax
\EndOfBibitem
\bibitem[Knupfer(2003)]{Knupfer2003}
Knupfer,~M. Exciton Binding Energies in Organic Semiconductors. \emph{Appl.
  Phys. A} \textbf{2003}, \emph{77}, 623--626\relax
\mciteBstWouldAddEndPuncttrue
\mciteSetBstMidEndSepPunct{\mcitedefaultmidpunct}
{\mcitedefaultendpunct}{\mcitedefaultseppunct}\relax
\EndOfBibitem
\bibitem[Yang \latin{et~al.}(2008)Yang, Shipman, Garrett-Roe, Johns, Strader,
  Szymanski, Muller, and Harris]{Yang2008}
Yang,~A.; Shipman,~S.~T.; Garrett-Roe,~S.; Johns,~J.; Strader,~M.;
  Szymanski,~P.; Muller,~E.; Harris,~C.~B. Two-Photon Photoemission of
  Ultrathin Film PTCDA Morphologies on Ag(111). \emph{J. Phys. Chem. C}
  \textbf{2008}, \emph{112}, 2506\relax
\mciteBstWouldAddEndPuncttrue
\mciteSetBstMidEndSepPunct{\mcitedefaultmidpunct}
{\mcitedefaultendpunct}{\mcitedefaultseppunct}\relax
\EndOfBibitem
\bibitem[Dutton \latin{et~al.}(2005)Dutton, Quinn, Lindstrom, and
  Zhu]{Dutton2005}
Dutton,~G.; Quinn,~D.~P.; Lindstrom,~C.~D.; Zhu,~X.-Y. Exciton Dynamics at
  Molecule-Metal Interfaces: C60 /Au(111). \emph{Phys. Rev. B} \textbf{2005},
  \emph{72}, 045441\relax
\mciteBstWouldAddEndPuncttrue
\mciteSetBstMidEndSepPunct{\mcitedefaultmidpunct}
{\mcitedefaultendpunct}{\mcitedefaultseppunct}\relax
\EndOfBibitem
\bibitem[Hagen \latin{et~al.}(2010)Hagen, Luo, Haag, Wolf, and
  Tegeder]{Hagen2010}
Hagen,~S.; Luo,~Y.; Haag,~R.; Wolf,~M.; Tegeder,~P. Electronic Structure and
  Electron Dynamics at an Organic Molecule/Metal Interface: Interface States of
  tetra-tert-butyl-Imine/Au(111). \emph{New J. Phys.} \textbf{2010}, \emph{12},
  125022 (17pp)\relax
\mciteBstWouldAddEndPuncttrue
\mciteSetBstMidEndSepPunct{\mcitedefaultmidpunct}
{\mcitedefaultendpunct}{\mcitedefaultseppunct}\relax
\EndOfBibitem
\bibitem[Varene \latin{et~al.}(2012)Varene, Bogner, Bronner, and
  Tegeder]{Varene2012}
Varene,~E.; Bogner,~L.; Bronner,~C.; Tegeder,~P. Ultrafast Exciton Population,
  Relaxation, and Decay Dynamics in Thin Oligothiophene Films. \emph{Phys. Rev.
  Lett.} \textbf{2012}, \emph{109}, 7601--1\relax
\mciteBstWouldAddEndPuncttrue
\mciteSetBstMidEndSepPunct{\mcitedefaultmidpunct}
{\mcitedefaultendpunct}{\mcitedefaultseppunct}\relax
\EndOfBibitem
\bibitem[Marks \latin{et~al.}(2013)Marks, Sachs, Schwalb, Sch\"{o}ll, and
  H\"{o}fer]{Marks2013}
Marks,~M.; Sachs,~S.; Schwalb,~C.; Sch\"{o}ll,~A.; H\"{o}fer,~U. Electronic
  Structure and Excited State Dynamics in Optcally Excited PTCDA Films
  Investigated with Two-Photon Photoemission. \emph{J. Chem. Phys.}
  \textbf{2013}, \emph{139}, 124701\relax
\mciteBstWouldAddEndPuncttrue
\mciteSetBstMidEndSepPunct{\mcitedefaultmidpunct}
{\mcitedefaultendpunct}{\mcitedefaultseppunct}\relax
\EndOfBibitem
\bibitem[Engel \latin{et~al.}(2005)Engel, Koschorreck, Leo, and
  Hoffmann]{Engel2005}
Engel,~E.; Koschorreck,~M.; Leo,~K.; Hoffmann,~M. Ultrafast Relaxation in
  Quasi-One-Dimensional Organic Molecular Crystals. \emph{Phys. Rev. Lett.}
  \textbf{2005}, \emph{95}, 157403\relax
\mciteBstWouldAddEndPuncttrue
\mciteSetBstMidEndSepPunct{\mcitedefaultmidpunct}
{\mcitedefaultendpunct}{\mcitedefaultseppunct}\relax
\EndOfBibitem
\bibitem[Guo \latin{et~al.}(2009)Guo, Ohkita, Benten, and Ito]{Guo2009}
Guo,~J.; Ohkita,~H.; Benten,~H.; Ito,~S. Near-IR Femtosecond Transient
  Absorption Spectroscopy of Ultrafast Polaron and Triplet Exciton Formation in
  Polythiophene Films with Different Regioregularities. \emph{J. Am. Chem.
  Soc.} \textbf{2009}, \emph{131}, 16869--16880\relax
\mciteBstWouldAddEndPuncttrue
\mciteSetBstMidEndSepPunct{\mcitedefaultmidpunct}
{\mcitedefaultendpunct}{\mcitedefaultseppunct}\relax
\EndOfBibitem
\bibitem[Chen \latin{et~al.}(2013)Chen, Barker, Reish, Gordon, and
  Hodgkiss]{Chen2013}
Chen,~K.; Barker,~A.; Reish,~M.; Gordon,~K.; Hodgkiss,~J.~M. Broadband
  Ultrafast Photoluminescence Spectroscopy Resolves Charge Photogeneration via
  Delocalized Hot Excitons in Polymer: Fullerene Photovoltaic Blends. \emph{J.
  Am. Chem. Soc.} \textbf{2013}, \emph{135}, 18502--18512\relax
\mciteBstWouldAddEndPuncttrue
\mciteSetBstMidEndSepPunct{\mcitedefaultmidpunct}
{\mcitedefaultendpunct}{\mcitedefaultseppunct}\relax
\EndOfBibitem
\bibitem[Tegeder(2012)]{Tegeder2012}
Tegeder,~P. Optically and Thermally Induced Molecular Switching Processes at
  Metal Surfaces. \emph{J. Phys.: Condens. Matter} \textbf{2012}, \emph{24},
  394001 (34pp)\relax
\mciteBstWouldAddEndPuncttrue
\mciteSetBstMidEndSepPunct{\mcitedefaultmidpunct}
{\mcitedefaultendpunct}{\mcitedefaultseppunct}\relax
\EndOfBibitem
\end{mcitethebibliography}

\end{document}